\documentclass[english]{iopart}
\usepackage[T1]{fontenc}
\usepackage[latin1]{inputenc}
\usepackage{float}
\usepackage{graphicx}

\makeatletter

\newcommand{\noun}[1]{\textsc{#1}}
\providecommand{\tabularnewline}{\\}

\usepackage{babel}
\makeatother

\begin{document}

\title[Microcalcification thickness determination]{The influence of scattered photons on the accurate determination
of microcalcification thickness in digital mammography }

\author{Varlen Grabski and Maria-Ester Brandan}

\address{Instituto de Fisica, UNAM, A. P. 20-364, 01000, D.F., Mexico}
\eads{\mailto{grabski@fisica.unam.mx}, \mailto{varlen.grabski@cern.ch}}

\begin{abstract}
Our interest has been to study the effect that scattered radiation
has on contrast , signal-to-noise ratio and thickness reconstruction
in digital mammographies. Using the GEANT code we have performed Monte-Carlo
simulations of 25 kVp Mo/Mo photons, through a breast phantom which
contains a 0.2-1.0 mm thick microcalcifications incident on a 20x106
$mm^{2}$ pixelized detector. The data have been analyzed assuming
6 different shapes of the incident beam: a 0.2x0.2 $mm^{2}$ {}``narrow''
beam, 4 different 20 mm long scanning beams of various widths, and
a 20x100 $mm^{2}$ beam with no scatter reduction mechanisms $(NSR)$
. Since the image of a point depends on scattered photons which passed
up to 2 cm away from the object (for 4 cm thick phantom), we identify
the background definition as a main source of systematic uncertainty
in the image quality analysis. We propose the use of two dimensional
functions (a polynomial for the background and Gaussians for the signal)
for total photon transmission description. Our main results indicate
the possible calcification thickness reconstruction with an accuracy
of the order of 6\% using 3 mm wide scanning beam. Signal-to-noise
ratio with the 3 mm wide beam gets improved by 20\% with respect to
$NSR$, a figure similar to that obtained with the narrow beam. Thickness
reconstruction is shown to be an alternative to signal-to-noise ratio
for microcalcification detection.
\end{abstract}

\submitto{PBM}

\maketitle

\section*{1. Introduction}

One of the main limitations of image reconstruction in mammography,
independent of external geometry and total breast thickness, is the
influence of scattered photons. The main parameter used to describe
scattered photon contributions is scatter-to primary radiation ratio
(SPR), which has been measured\cite{barnes_b} and calculated \cite{boon_coop},
and is $\sim0.5-0.6$ for typical voltages and breast phantom dimensions.
The presence at the image receptor \emph{of} photons which have been
scattered by the breast tissue components results in a severe loss
of contrast. The contrast reduction is estimated to be about $\sim$0.6
for the above SPR, and the range of values encountered in mammography
indicates that the contrast can be improved by factors of $\sim$1.4
- 2.5 if scattered radiation is eliminated from the image \cite{Barn_sc}.
The most popular solution for the problem has been the use of antiscatter
grids in mammography units \cite{Rezents}. These improve the contrast
by typical factors of $\sim$1.2 - 1.4 but also reduce the primary
intensity, resulting in a patient dose increase of approximately 2
for typical mammographic conditions \cite{Rezents} in order to reach
the necessary photon fluence for a good quality image. Lately, the
use of scanning narrow beams \cite{Barn_sslit,Yester}, a direct way
to reduce the volume of the irradiated scattering medium, has found
its way into commercial mammographic systems.

Mammographic systems using digital detectors offer distinct advantages
with respect to the conventional screen/film image recorder due to
\textbf{}their much wider dynamical range of \textbf{}useful \textbf{\noun{}}exposures.
Since the response of the detector is linear over some 4 orders of
magnitude in exposure, \textbf{}there is no need to increase exposure
if scattering reduction methods are used, and the signal detection
limit is dominated by the signal-to-noise ratio ($SNR$) \cite{Neitzel}.
Another parameter which can be used to evaluate the digital image
quality is the linear size of the object along the photon direction,
as evaluated from the image. This thickness can be determined by using
the visual contrast \textbf{}(VC) \cite{David} and the linear absorption
coefficients. Clinically, the correct determination of this parameter,
for instance, a microcalcification $(\mu C)$ thickness, can be an
indicator of the stage of development of this formation. \textbf{}Though
we consider the thickness parameter to be of more evident and real
character than the contrast, it is necessary to determine which of
these two is more sensitive, both for the micro-calcification detection
and its study as well.

In this work we have performed Monte Carlo (M-C) simulations of the
passing of photons through a breast phantom which contains a few $\mu C$
of different thickness inside. The influence of the scattered photons
on the contrast, $SNR$ and thickness determination has been studied
for a variety of incident beam definitions, using a scanning slit
as the scatter reduction technique.

\section*{2. The model }

At least two X-ray transmission measurements are necessary to recover
each of the components of a simulated three-component breast (microcalcification,
adipose and glandular tissue), if the total thickness is known\cite{Lemack}.
This task can be considerably simplified under the assumption that
one of the components has constant thickness; this will lead to a
two-component model, for which one measurement is enough. The phantom
proposed in this work assumes that microcalcifications are embedded
in the glandular tissue, and that constant-thickness adipose tissue
covers the outside of the breast. The geometry and the structure of
the proposed breast phantom model is shown in Fig. 1. The phantom,
with lateral dimensions $10x10$ $cm^{2}$, has a total thickness
of $4$ $cm$. Two adipose layers cover the top and bottom sides with
a total thickness of 1 cm, and 3 cm thick glandular tissue is located
between the adipose layers. Five $\mu C$ are located at the midplane
of the glandular tissue layer. The microcalcifications are cylindrical
in shape, 4 mm diameter, and have variable thicknesses between $0.2$
and $1.0$ $mm$. The selection of a $25$ percent adipose component
is done for the purpose of increasing the relative contribution of
the noise\cite{Lemack}. The detailed chemical compositions of the
phantom materials are presented in table 1. All estimates, and the
M-C simulations, have been carried out with this simplified model
of the breast. 

In order to understand the difficulties to recover the $\mu C$ dimensions
from a radiological image, let us conduct some estimations for mono-energetic
photons, neglecting the effects of scattering. In the absence of $\mu C$,
the number of photons passing through the phantom $(N_{nc}(x,y))$
is defined through the total number of $N_{0}(x,y)$ primary photons
as:

\begin{eqnarray}
N_{nc}(x,y) & = & N_{0}(x,y)\exp(-\mu_{a}t_{a}(x,y)-\mu{}_{g}t_{g}(x,y)),
\label{eq:1}\end{eqnarray}

where $\mu_{a}$ and $\mu_{g}$ are the linear absorption coefficients,
and $t_{a}$ and $t_{g}$ are the thickness of the adipose and glandular
tissues, respectively. With the addition of $\mu C$, Eqn. (1) transforms
into: 
\begin{eqnarray}
N_{c}(x,y) & = & N_{0}(x,y)\exp(-\mu_{a}t_{a}(x,y)-\mu{}_{g}t_{g}^{c}(x,y)-\mu_{c}t_{c}(x,y)),
\label{eq:2}\end{eqnarray}

where $\mu_{c}$ is the microcalcification linear attenuation coefficient,
and $t_{c}(x,y)$ is its thickness. $N_{c}(x,y)$ is the number of
the transmitted photons in the presence of calcifications. Within
our simplified model, $t_{g}^{c}(x,y)$ will be defined as:

\begin{eqnarray}
t_{g}^{c}(x,y) & = & t_{g}(x,y)-t_{c}(x,y).\label{eq:3}
\end{eqnarray}

Dividing Eqn. (1) into (2) and taking logarithms, we obtain the following
for the $\mu C$ thickness $t_{c}$:

\begin{eqnarray}
t_{c}(x,y) & = & D_{\mu}^{-1}\log(N_{nc}(x,y)/N_{c}(x,y)),\label{eq:4}
\end{eqnarray}

where $D_{\mu}=\mu_{c}-\mu_{g}$. To be correct, linear attenuation
coefficients such as those in the NIST data base, should be used only
for the narrow-beam condition\cite{weight} since they do not include
the effect of the scatter radiation\textbf{.} Thus, within this approximation,
the $\mu C$ thickness can be easily determined by one measurement
in which the value of $N_{nc}$ is determined from the region outside
the microcalcification. However, in reality, the determination of
$N_{nc}$ is only approximate because the effects of scattering, geometry
and the inner structure of the breast tissue can introduce several
inaccuracies. 

The contrast parameter, traditionally used in conventional mammography,
is useless in digital mammography since the possibility of detecting
the signal depends on the $SNR$\cite{Neitzel}, defined as:

\begin{eqnarray}
SNR & = & (N_{nc}-N_{c})/\sqrt{N_{nc}+N_{c}}.
\label{eq:5}
\end{eqnarray}

The important question is the choice of the most appropriate parameter
to use for the accomplishment of the image quality optimization. \textbf{}

Let us compare $SNR$ and $t_{c}/\sigma_{t_{c}}$ to determine which
one is more sensitive for the detection of the $\mu C$. The ratio
$SNR/(t_{c}/\sigma_{t_{c}})$ can be written as:

\begin{eqnarray}
SNR/(t_{c}/\sigma_{t_{c}}) & = & \frac{1-m}{\sqrt{m}\cdot\log(1/m)},\label{eq:6}\end{eqnarray}

where $\sigma_{t_{c}}$ is the uncertainties in the determination
of thickness, $t_{c}$ and $SNR$ are given by Eq 4 and 5, respectively,
and $m=N_{c}/N_{nc}$ . When $m\rightarrow1$, which corresponds to
a thin $\mu C$, relation (6) approaches 1, which indicates the equivalence
of $t_{c}/\sigma_{t{}_{c}}$ and $SNR$ parameters for the detection
of the $\mu C$, and the relative uncertainties of both parameters
have identical statistical behavior, $\sim1/\sqrt{N_{nc}}$.On the other hand,
in case of a given fluence,the statistics are proportional to the object 
surface so the relative statistical uncertainty in the object image will
depend on $1/r$, where $r$ is the object linear size. Consequently, 
the measurement of the $\mu C$ thickness is not less sensitive than measuring
the $SNR$ or contrast in the detection of the $\mu C$ and, at the same
time, makes possible to restore the $\mu C$ three dimensions. 

Everything stated above is correct in the absence of scattering. After
switching scattering on, the description becomes more complicated
and conducting estimations is complex and dependent on the geometry
and structure of the breast. This problem can be easily solved using
a simulation of the photon transport through the phantom volume.

\section*{3. Monte-Carlo simulation}

There are two different possibilities for  M-C simulation
of the photon transport process in the phantom. The results of M-C
simulations based on the convolution method \cite{boon_coop} (also
known as {}``fast'' M-C simulation), are sensitive to geometry and
beam parameters. This is why it is necessary to estimate the possible
systematic uncertainties of the method each time it is used when geometry,
medium, beam size, etc are changed. The method that we use in this
work ({}``full'' simulation) \textbf{}is based on the individual
transport of each photon. It is not as fast as convolution, but is
more accurate when describing the concrete experimental conditions.
The choice of the method depends on the task. In our opinion, the
code GEANT\cite{geant} is a very good choice for this purpose. This
powerful Monte Carlo program was built for the transport of elementary
particles through matter, and includes all processes of low energy
photon interactions which are relevant for the transport of typical
mammography X-rays. This program, which has been tested to be appropriate
for the high-energy region, it is now more and more frequently used
in medical physics\cite{medgeant}. GEANT4 is flexible enough for
the required additional programming in C++, and is user-friendly.
To reconstruct the thicknesses from the photon intensities at the
image detector plane we will use mass attenuation coefficients from
the NIST data base \cite{NIST}. The mass attenuation coefficients
for tissue components and calcium carbonate are calculated using percentages
per weight according to \cite{weight} and NIST and shown in Table
1. Our estimates indicate that the agreement between these mass coefficients
data and GEANT internal cross sections for the physical processes
is not worse than 2 percent in the energy region below 25 keV. At
this stage, this agreement is sufficient to study the influence of
scattering on the accuracy of the thickness determination.  To use
the code it is necessary to describe the photon beams incident on
the phantom and the geometry and composition of the detector. The
simulated experimental setup is shown in Fig. 2. Almost all significant
characteristics of the digital mammography unit Senographe 2000D (GE
Medical Systems) have been incorporated in this setup with the purpose
to assume parameters of an existing system. No antiscatter grid is
being used. For the photon beam, a typical X-ray spectrum for 25 kVp
Mo/Mo target/filter combination has been used \textbf{}\cite{handbook}.
The angular distribution of X-rays on the phantom has been assumed
uniform. Photons have been detected by pixelized 0.1x 0.1x
0.1 $mm^{3}$ CsI(Tl) scintillators covering a total area equal 20x106
$mm^{2}$. The beam size on phantom was 20x106 $mm^{2}$, equal to
the detector size, with the purpose of decreasing the simulation time.
The total number of primary photons incident on the $\sim21$ $cm^{2}$
phantom surface is $\sim1.8x10^{9}$, which corresponds to a normalized
glandular dose $\sim0.03$ $mGy$ \cite{Boone_alone}. This dose is
rather low compared with the usual values in mammography\textbf{.}
The results of the simulation have been stored in binary files for
the offline analysis, performed by a program, that uses the mathematical
and graphic library ROOT\cite{root}.

\section*{4. Results and discussion}

The simulated data were analyzed assuming 6 different shapes of the
incident beam. All these \textbf{}beams have rectangular shapes on
the phantom. The {}``ideal'' beam is narrow with dimensions $0.2x0.2$
$mm^{2}$. For the other beams, one dimension is always equal to the
width of the detector (20 mm) and the other is variable. We use the
following nomenclature:

$0.2x0.2mm^{2}$, \textbf{{}``}narrow'' \textbf{}beam\textbf{;}

$1x20mm^{2}$ , 1mm wide scanning beam ;

$3x20mm^{2}$, 3mm wide scanning beam;

$5x20mm^{2}$ , 5mm wide scanning beam ;

$10x20mm^{2}$, 10mm wide scanning beam ;

$20x100mm^{2}$ ,{}``NSR'' , non-scattering reduction.

Data scanning was done along the detector long axis and beam size
was controlled by the collimator placed between the X-ray source and
the phantom. For the simulation of the scanning beam, the analysis
included only the data generated by photons incident within a collimator
region. The NSR regime didn't use any scan.

In order to determine the characteristic size of the region of scattered
photons, in Fig 3 we have plotted the distribution of scattered photons,
point spread function (PSF), as a function of Dx (coordinate difference
between the initial and the scattered photons position ). The spot
size defined as the PSF (root-mean-squared ) is $\sim1$ cm. For our
geometry the SPR is 0.39, which agrees with similar M-C \cite{boon_coop}
calculations for 4 cm thick phantoms and 25 kVp X-rays. The distribution
in Fig. 3 shows that the image of each point depends on photons that
pass up to 2 cm away from the point . This value depends on the geometry
and will increase as a function of total phantom thickness. This result
also indicates that, in order to determine the value of $N_{nc}$,
it is necessary to define a distance more than 2 cm away from the
$\mu C$. But, this distance is sufficiently large for the structure
and geometry of the phantom to have changed. That's why we suggest
a different procedure of background calculation. 

The total signal $F(x,y)$ in the image (distribution of photons on
the detector) can be expressed as the sum of the $\mu C$ and the
background signals, where the background $P(x,y)$ is supposed to
show smooth behavior and the $\mu C$ signal $G(x,y)$ is described
using a Gaussian function:

\begin{eqnarray}
F(x,y) & = & P(x,y)+\Sigma G(x,y),\label{eq:7}\end{eqnarray}

where $P(x,y)$ is a two-dimensional polynomial of order three and
$G(x,y)$ is a two-dimensional Gaussian function for each target $\mu C$.
The parameters of this function have been defined by fits on simulation
$N_{c}(x,y)$ data. The number of parameters in $F(x,y)$ is 25 and
the number of points $\sim$ 8000. The value of $\chi^{2}$ per point
is typically 1.5 - 2.5 which is not bad (taking into account the approximate
description of the signal with Gaussian shapes). The description of
the target images as having Gaussian shapes may not be the best, but
it makes the task easier. After defining its parameters by fit, the
function $P(x,y)$ has been used as the background instead of $N_{nc}(x,y)$
in the $SNR$, contrast and thickness definitions. A symmetric noise
in the thickness, $SNR$ and contrast definitions with respect to
zero, indicates that the fit is appropriate. 

The $F(x,y)$ for all the events (20x106 $mm^{2}$ X-ray beam ) is
plotted in Fig 4. The decreasing values of $F(x,y)$ near the edges
of the detector can be explained as a geometrical and scattering effect.
By using an extreme scatter reduction method (0.2x0.2 $mm^{2}$ X-ray
beam ), as shown in Fig. 5, it is possible to make the background
behavior more flat ($\pm1$\% compared with a plane surface). To reduce
the number of parameters, the standard deviations $\sigma_{x}$ and
$\sigma_{y}$ for each Gaussian-shaped $\mu C$ have been set equal
( $\sigma_{x}=\sigma_{y}$ ). The calculated diameters for all $\mu C$
are plotted in Fig 6. Error bars are parameter errors obtained during
the fit. The overestimations of the transversal sizes (diameter) ($\sim$
25\%) can be explained as the consequence of a not-totally appropriate
description of the signal by the Gaussian functions. The thickness
dependence of the (calculated /original) $\mu C$ diameter ratio in
Fig. 6 can be explained as the increase of scattering as a function
of the $\mu C$ thickness. 

The $SNR$ is defined as:

\begin{eqnarray}
SNR(x,y) & = & (P(x,y)-N_{c}(x,y))/\sqrt{(\sigma_{P(x,y)})^{2}+N_{c}(x,y)},
\label{eq:8}\end{eqnarray}

where $\sigma_{P(x,y)}$ is the definition uncertainty of the $P(x,y)$,
which should be smaller than $\sqrt{P(x,y)}$. In the calculations
we have used the value $\sqrt{P(x,y)}$ for $\sigma_{P(x,y)}$.

The contrast C is defined as:

\begin{eqnarray}
C(x,y) & = & (P(x,y)-N_{c}(x,y))/P(x,y).\label{eq:9}\end{eqnarray}

The thickness $t_{c}$ is defined as:

\begin{eqnarray}
t_{c}(x,y) & = & D_{\mu}^{-1}\log(P(x,y)/N_{c}(x,y))\label{eq:10}\end{eqnarray}

and, from the fits, $t_{f}$ is defined as:

\begin{eqnarray}
t_{f}(x,y) & = & D_{\mu}\log(P(x,y)/F(x,y))\label{eq:11}\end{eqnarray}

where $N_{c}(x,y)$ is the number of photons detected in the pixel
detector, and $D_{\mu}$ is the difference between mean values of
the linear attenuation coefficients for $\mu C$ and glandular tissue.
The results for $t_{f}$ are shown in Fig 7 as the ratio of the reconstructed
( Gaussian maximum for each target) value from Eqn.(11) to the original
thicknesses, for the {}``narrow'' beam. For the other beams, we
show their ratio with respect to the {}``narrow''. In order to get
the correct thickness for the 200 $\mu m$ calcifications it is necessary
to make a correction due the peaked value of the Gaussian shape (suggested
correction: a 0.73 factor \textbf{}on signal value $(P(x,y)-F(x,y))$
for the 200 $\mu m$ and slowly increasing up to 0.80 for 1000 $\mu m$).
As we have mentioned before, the Gaussian fit is not optimum for these
cylindrical target shapes (diameter/thickness ratio is $\sim$4 -
20) because the flat tops are noticeable in the image. 

Results of the 3-dimensional reconstructed distributions of $t_{c}(x,y)$,
$SNR(x,y)$ and $C(x,y)$ for our phantom are shown in Figs. 8 - 10.
These distributions have been used to determine the mean values of
the thickness, $SNR$ and contrast of the targets. To reduce statistical
errors we have calculated the mean values of the thickness only for
the cases where they are greater than the original thickness minus
3 noise values. The central target is 200 $\mu m$ thick and the collected
statistics is enough to have a signal 10 standard deviations above
the noise for the 'narrow' beam (see Fig 8). For the 200 $\mu m$
thick $\mu C$, the signal to noise ratio is $\sim$ 10, (Fig 9) which
agrees with our statement that thickness and $SNR$ have similar sensitivities
for $\mu C$ detection. The bin size of the histograms is
0.5x0.5 $mm^{2}$. The same level of  statistical errors
for the detector pixel size (0.1x0.1 $mm^{2}$) can be reached increasing the
dose approximately 25 times that is 0.75 mG, which, is still low compared with usual dose level.

The main source of systematic uncertainty in the thickness, $SNR$
or contrast definitions (shown in Figs 11-13) for cases with and without
reduction of scattering, is the uncertainty in the background definition.
The background definition could be improved using a better signal
description\textbf{.} So, it is necessary to use functions with more
parameters for a better signal description as well as scatter reduction
methods to improve the definitions of the above mentioned parameters.

\subsection*{4.1 $SNR$ and Contrast}

Contrast calculations have been done only to compare with other calculations
and experimental data. The $SNR$ and contrast in the simulated images
for different $\mu C$ thicknesses and for different beams are shown
in Figs 12 and 13. No appreciable difference is observed between the
$SNR$ and contrast dependences on the $\mu C$ thicknesses. The $SNR$
dependence is more linear than the contrast and less sensitive to
the scattered photon contribution. It is evident from Figs 12 and
13 that it is possible to improve $SNR$ and contrast by approximately
$\sim17,28$\%, respectively, using the {}``narrow'' beam . This
advantage is almost independent of the $\mu C$ thickness up to the
1 mm region. Both parameters improve when the scanning beam size decreases.
It seems possible to reach a $\sim$17-27\% improvement for the 3mm
wide beam.

Results on contrast improvement using grid or scanning beam that we
find in the literature are diverse. Differences in experimental data
are large, going from no improvement at all \cite{Weldkamp} up to
50\% \cite{boonerad,Rezents}. The M-C calculation in \cite{boon_coop}
predicts a 40\% contrast improvement; the apparent discrepancy with
these results can be explained from differences in geometry. In our
case, one dimension of the scanning beam is limited to 2 cm and the
contribution of scattered photons doesn't reach its maximum possible
value. As can be seen in Fig. 4, the background value depends on the
distance of the beam to the phantom edge. Therefore, for the contrast
defined by Eqn. (9) the improvement depends on the coordinate and
will increase far away from the boundaries. 

In fact, image quality improvement can be explained using the statistical
properties if the statistical noise is dominant. For a given incident
exposure, for the low $SNR$ values close to the detection threshold
and with approximation signal $\ll$ background, the $SNR$ improvement
parameter $SI$ ( $SI=SNR^{with}/SNR^{without}$ , where $SNR^{with}$
and $SNR^{without}$ are the signal to noise ratios with and without
scatter reduction methods, respectively ) can be written as:

\begin{eqnarray}
SI & = & \frac{k_{p}\sqrt{1+SPR}}{\sqrt{k_{p}+k_{s}SPR}}=k_{p}\sqrt{BF},
\label{eq:12}\end{eqnarray}

where $k_{p}=N_{p}^{with}/N_{p}^{without}$, $k_{s}=N_{s}^{with}/N_{s}^{without}$,
are the transmission coefficients for primary and scattered photons,
respectively. $N_{p}^{with}$, $N_{p}^{without}$, $N_{s}^{with}$
and $N_{s}^{without}$ are the numbers of primary and scattered photons
with and without scatter reduction methods, respectively. BF is the
Bucky factor of the scatter reduction grids\cite{Barn_sslit,Fahirg_BF}.
So, in case of using grids for scatter reduction and in order to have
improvement in $SNR$ without any additional dose, it is necessary
to provide the following condition for the transmission coefficients:

\begin{eqnarray}
k_{p} & \geq & \frac{1+\sqrt{1+4k_{s}SPR(1+SPR)}}{2(1+SPR)}.
\label{eq:13}\end{eqnarray}

For scanning beams the primary photons transmission coefficients $k_{p}$
are always 1 and the improvement depends on $k_{s}$ and $SPR$. The
$SI$ maximum value only depends on the $SPR$ value, and is equal
to $SI_{max}=\sqrt{1+SPR}$. To obtain a given value of $SI\leq SI_{max}$,
the transmission coefficient of the scattered photons satisfies:

\begin{eqnarray}
k_{s} & = & \frac{1}{SI^{2}}(1-\frac{SI^{2}-1}{SPR})
\label{eq:14}\end{eqnarray}

For this transmission coefficient it is possible to calculate the
beam sizes using the point spread function (see Fig 3).

\subsection*{4.2 Thickness}

Results of the calculated $\mu C$ thickness for the different beams
are shown in Figs. 11 and 14. In Fig. 11 we show the thickness ratio
for different dimensions of the scanning beams. For the \char`\"{}narrow\char`\"{}
beam\textbf{,} the systematic uncertainty in the background definition
is less than 1\% (the flatness of the background for the {}``narrow''
beam is $\sim$ 1\%, see Fig 5). There is a $\sim$1-2\% systematic
discrepancy between the calculated and the original thicknesses for
the 200 $\mu m$ calcification, which increases up to $~3-4$\% for
the 1 mm thick $\mu C$. The $1-2$ \% disagreement could be related
to differences in the mass attenuation coefficients used by GEANT
during simulation and NIST data used for the thickness reconstruction.
This source of systematic uncertainty is important only for the absolute
thickness definitions and will be smaller for the thickness ratio
definitions. Another possible source for this systematic uncertainty
is the uncertainty in the calcium carbonate mass attenuation coefficient
calculation. 

The improvement of thickness determination when suppressing the scattering
is approximately 35\% for small thicknesses and increases up to 45
\% for a 1 mm thick calcification as shown by Fig 11. This improvement
can be greater when using wider beams , since the contribution of
scattered events can increase by 30\% (see Fig. 3) and this would
change the background by $\sim$ 10\%. The dependence of the thickness
improvement on the beam size in the scanned direction for the different
$\mu C$ thickness is shown in Fig 14. The improvement for all $\mu C$
thicknesses will decrease with beam size and for the 3mm wide beam
it will reach of 85-95 \% of the \char`\"{}narrow\char`\"{} beam.

We have not been able to find published experimental information about
the use of the thickness determination. There are data for visual
contrast improvement\cite{Rezents}. \textbf{}This \textbf{}concept
is equivalent to thickness improvement, since the attenuation coefficients
cancel out \textbf{\emph{}}in ratio. Data for VC \cite{Rezents} for
phantom sizes 12.4x12.4x4 $cm^{3}$ (50/50\% adipose /glandular) phantom
with different exposed beams and grid types show maximal VC improvement,
up to 50\% (uncertainty $\sim$ 5\%). Our results agree with these
data, taking into account geometrical and phantom differences.

\subsubsection*{4.2.1 $\mu C$ position uncertainty}

Everything mentioned above is correct when the $\mu C$ target positions
are known. \textbf{}Usually, the $\mu C$ positions in the breast
are unknown and calculations of the absorption coefficient averages
(see 4) are problematic, among other reasons, because photon spectrum
will strongly depend on the depth of the point in the phantom. \textbf{}In
our case, the \textbf{$\mu C$} absorption coefficient mean values
have been calculated assuming the photon spectrum at the center of
the \textbf{}phantom. The systematic uncertainty introduced by this
simplification may be estimated for the 25 kVp Mo/Mo spectrum and
4 cm thick phantom assumed in this study, as being $\sim$2.5\% in
$D_{\mu}$. This value could be decreased, at least twice, using additional
filters that narrow the energy spectrum. These systematic uncertainties
will increase with total phantom thickness. To make microcalcification
thickness measurements independent of the breast thicknesses it would
be necessary to use mono-energetic photon beams.

\section*{Conclusion}

We have preformed Monte-Carlo simulation of 25 kVp Mo/Mo X-rays transported
in a 4 cm thick breast phantom. We have focused on the reconstruction
of the thicknesses of 0.2-1.0 mm thick microcalcifications embedded
in the phantom. We have shown the possible thickness reconstruction
with an accuracy of the order of 6\% using a 3 mm wide slot scanning
beam. This slot size, which seems to be technicaly feasible, for a
mamography unit \textbf{}promises results which are close to the ideal
narrow beam. \textbf{}The same beam could improve the signal-to-noise
ratio by $\sim20$ \%, similar to the effect of using ideal narrow
beam. 

The $\mu C$ thicknesses parameter can be used as alternative to \textbf{$SNR$}
for microcalcification detection. \textbf{}The use of semi mono-energetic
photon beams \textbf{}would decrease systematic uncertainties in $\mu C$
thicknesses determination. One of the goals of this calculation was
to show that the GEANT code is appropriate for digital mammography
calculations.

\section*{Acknowledgments}

Authors thank partial support from DGAPA-UNAM, Grant IN-109302

\maketitle

\maketitle

\begin{table}[H]
\caption{Chemical composition by weight, of the phantom materials. Values
are taken from NIST\cite{NIST}}
\begin{tabular}{|c|c|c|c|}
\hline 
Z &
Adipose tissue&
Glandular tissue&
Microcalcification$(CaCO_{3})$\tabularnewline
\hline
\hline 
1&
0.114&
0.106&
-\tabularnewline
\hline 
6&
0.598&
0.332&
0.12\tabularnewline
\hline 
7&
0.007&
0.03&
-\tabularnewline
\hline 
8&
0.278&
0.527&
0.48\tabularnewline
\hline 
11&
0.001&
0.001&
-\tabularnewline
\hline 
15&
-&
0.001&
\tabularnewline
\hline 
16&
0.001&
0.002&
\tabularnewline
\hline 
17&
0.001&
0.001&
\tabularnewline
\hline 
20&
-&
-&
0.40\tabularnewline
\hline
\end{tabular}

\end{table}

\maketitle

\begin{figure}[H]
\includegraphics[%
  scale=0.8]{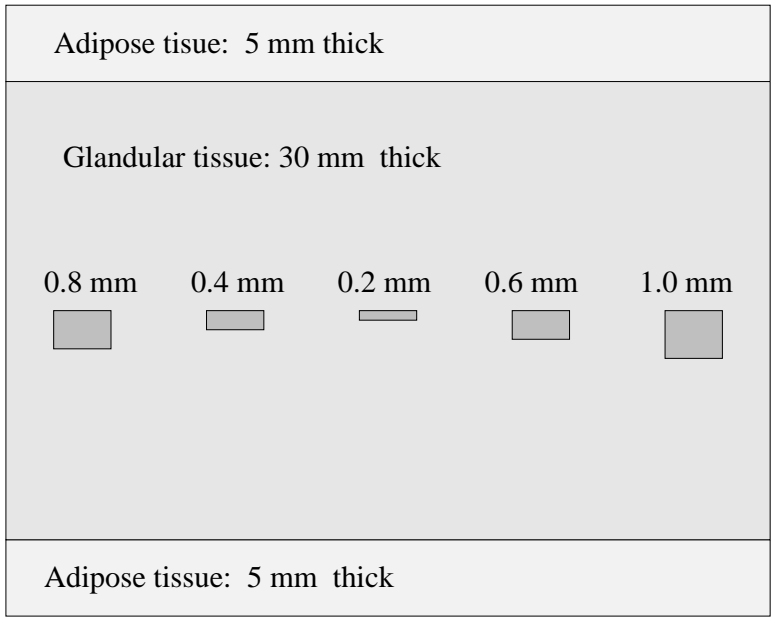}

\caption{Phantom structure. Calcification thickness is indicated. Cylindrical
microcalcifications are 4 mm in diameter.}
\end{figure}

\begin{figure}[H]
\includegraphics{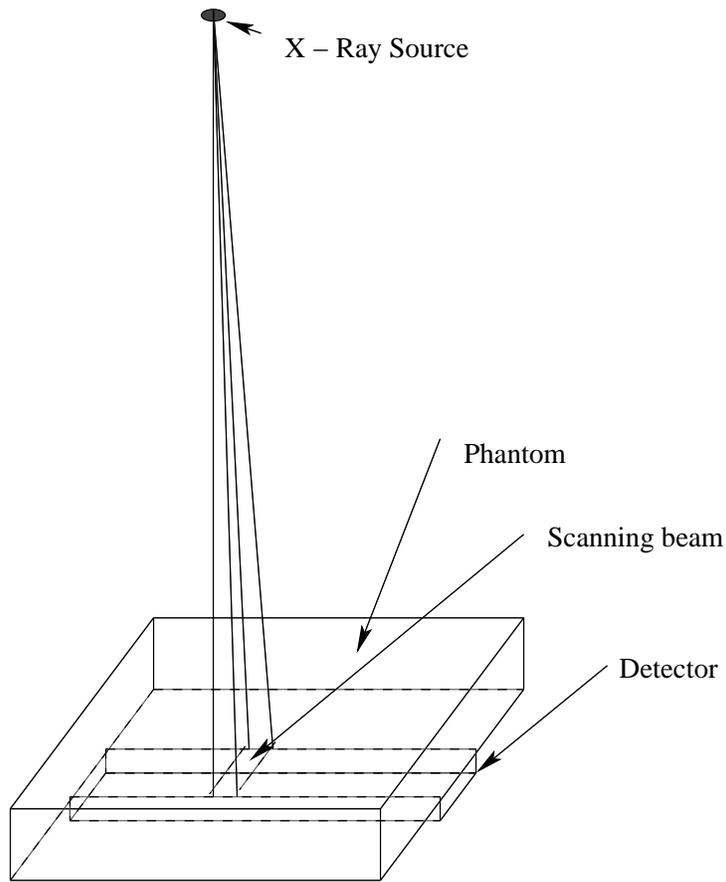}

\caption{Simulation setup}
\end{figure}

\begin{figure}[H]
\includegraphics[%
  scale=0.8]{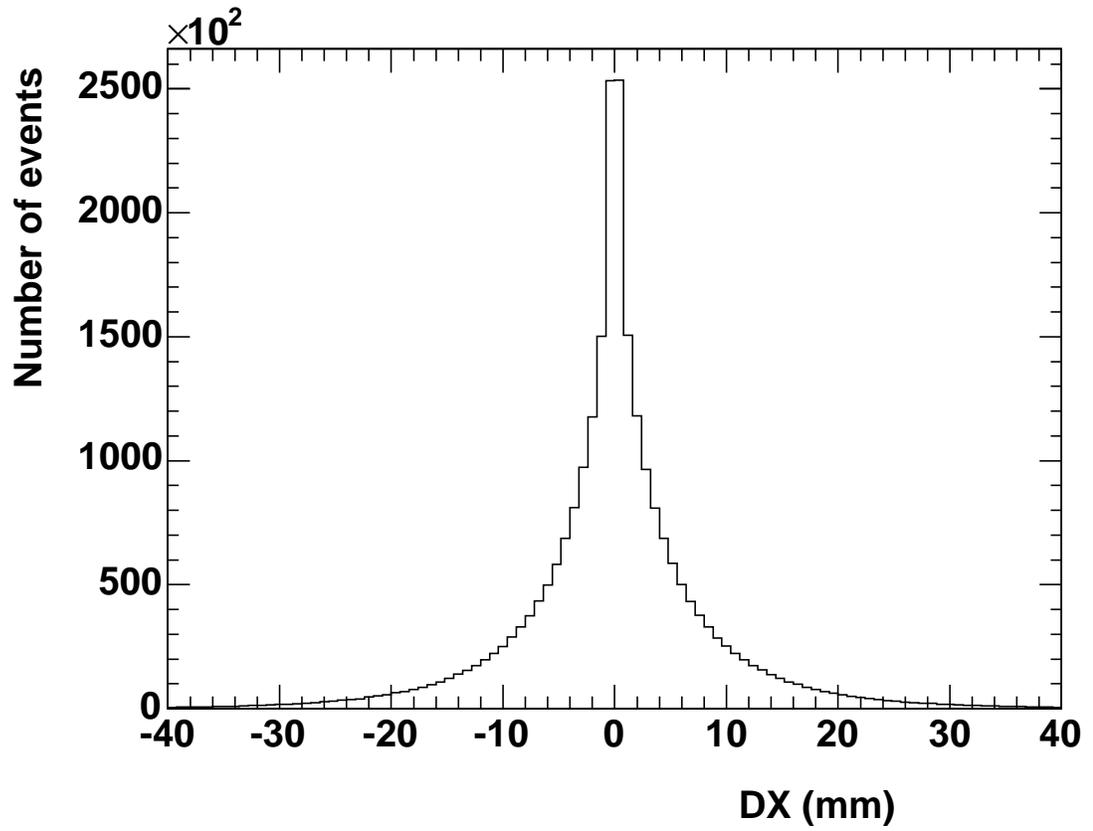}

\caption{Point spread function for scattered photons (see text)}
\end{figure}

\begin{figure}[H]
\includegraphics[%
  scale=0.8]{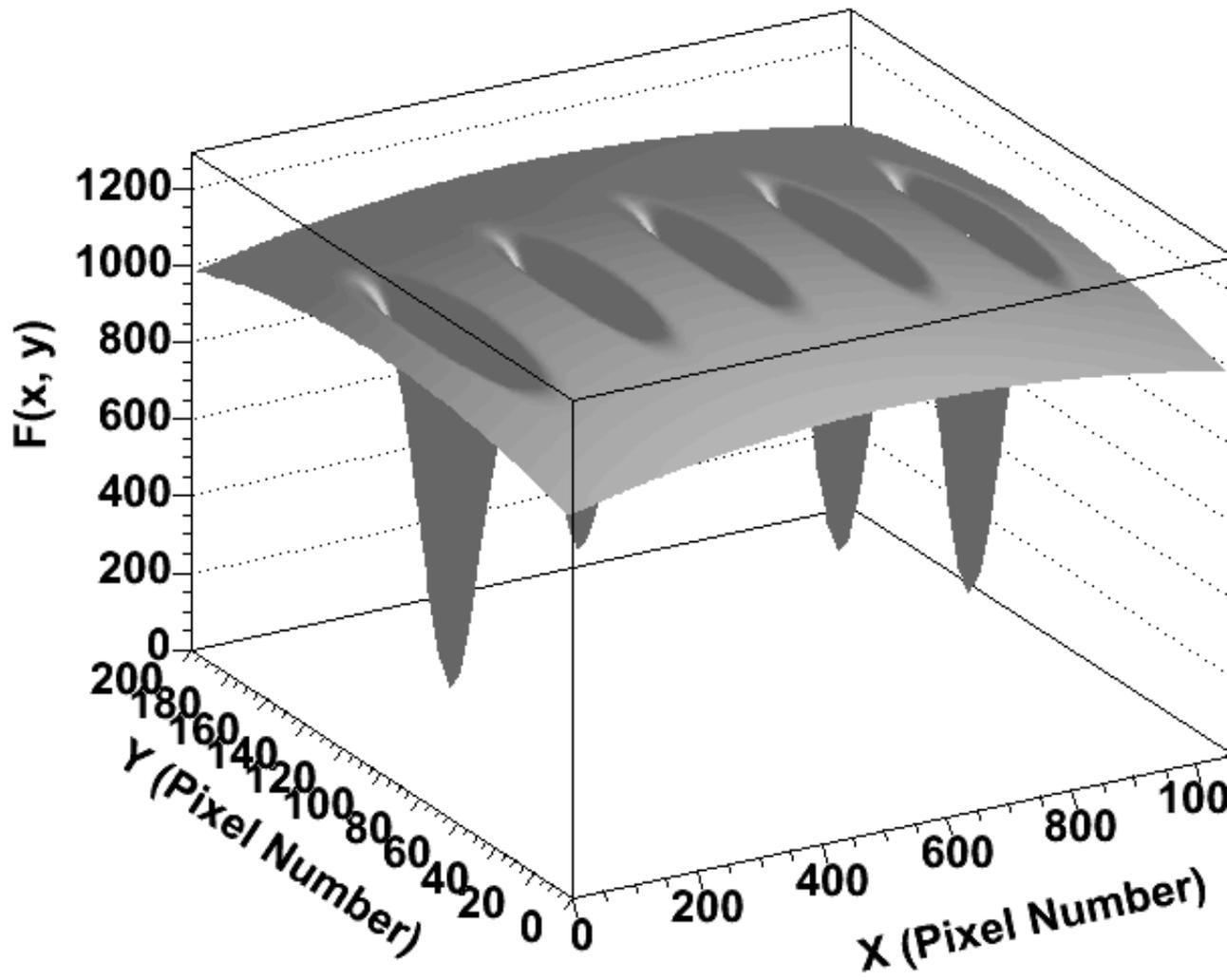}

\caption{Signal and background description by a multi-parameter function F(x,y)
in the case NSR, without scanning (see text)}
\end{figure}

\begin{figure}[H]
\includegraphics[%
  scale=0.8]{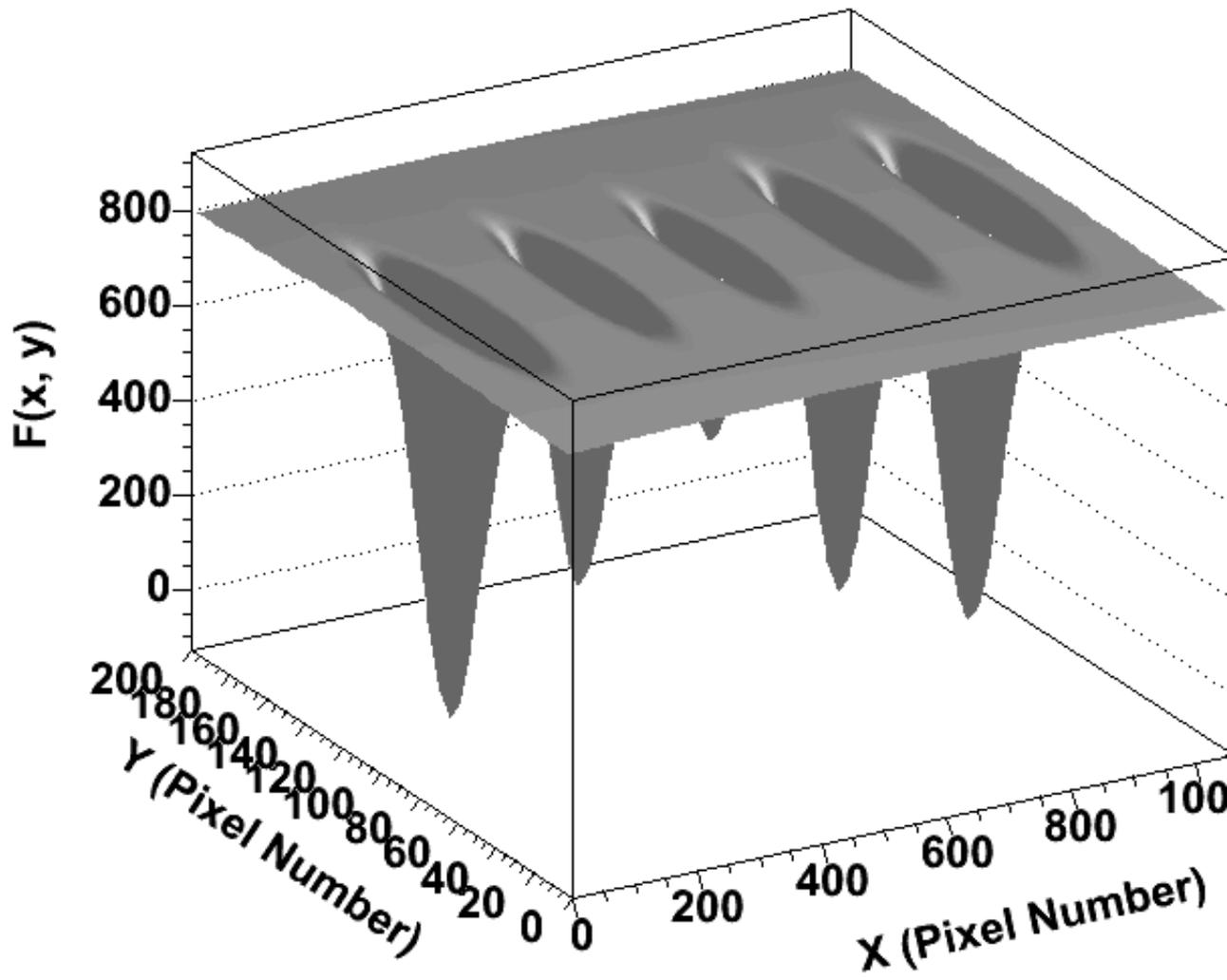}

\caption{The same as in Fig 4 for case of the {}``narrow'' beam (see text)}
\end{figure}

\begin{figure}[H]
\includegraphics[%
  scale=0.8]{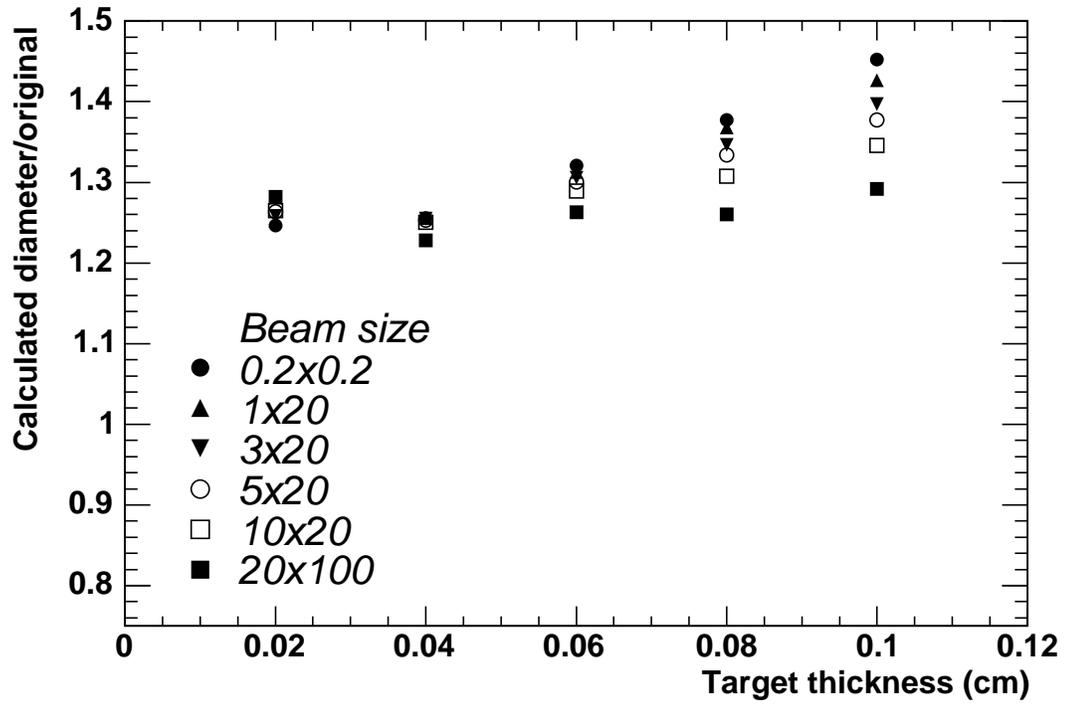}

\caption{Calculated $\mu C$ diameter $(\sigma_{\mu C}^{fit}\sqrt{3})$ as
a function of the micro-calcification thickness. Beam sizes are in
$mm^{2}$.}
\end{figure}

\begin{figure}[H]
\includegraphics[%
  scale=0.8]{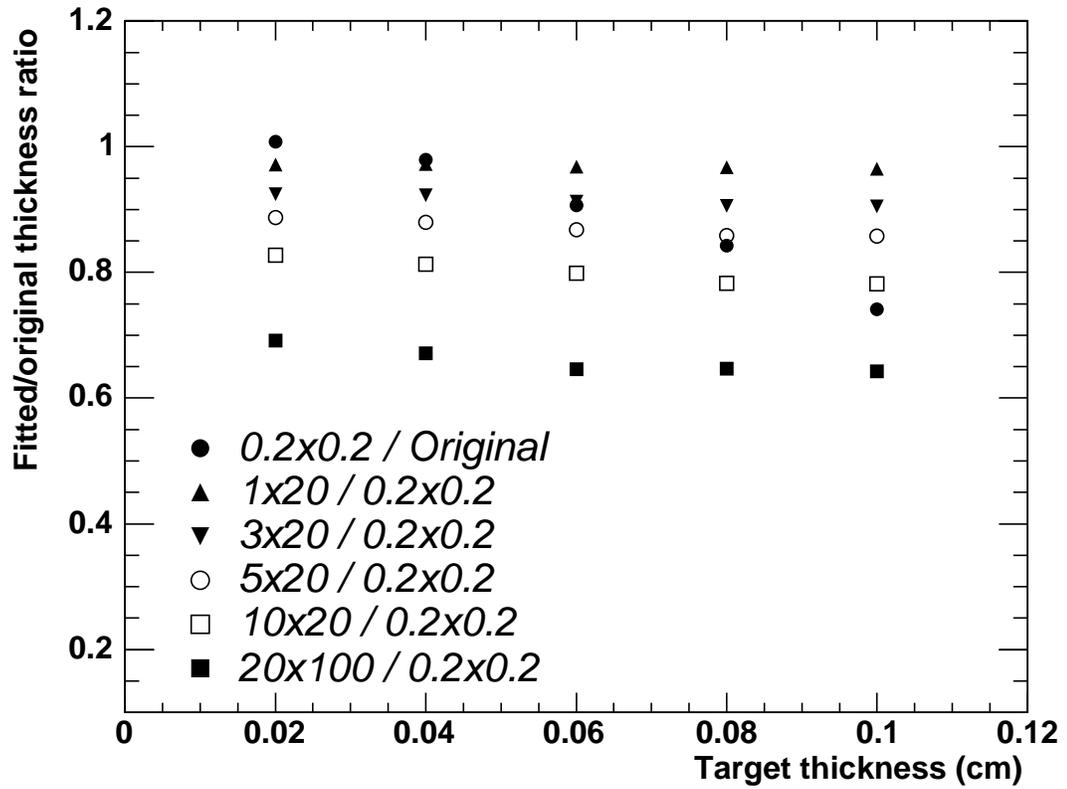}

\caption{Thickness calculation ($t_{f}$) using the function $F(x,y)$ (see
text). Beam sizes are in $mm^{2}$.}
\end{figure}

\begin{figure}[H]
\includegraphics[%
  scale=0.8]{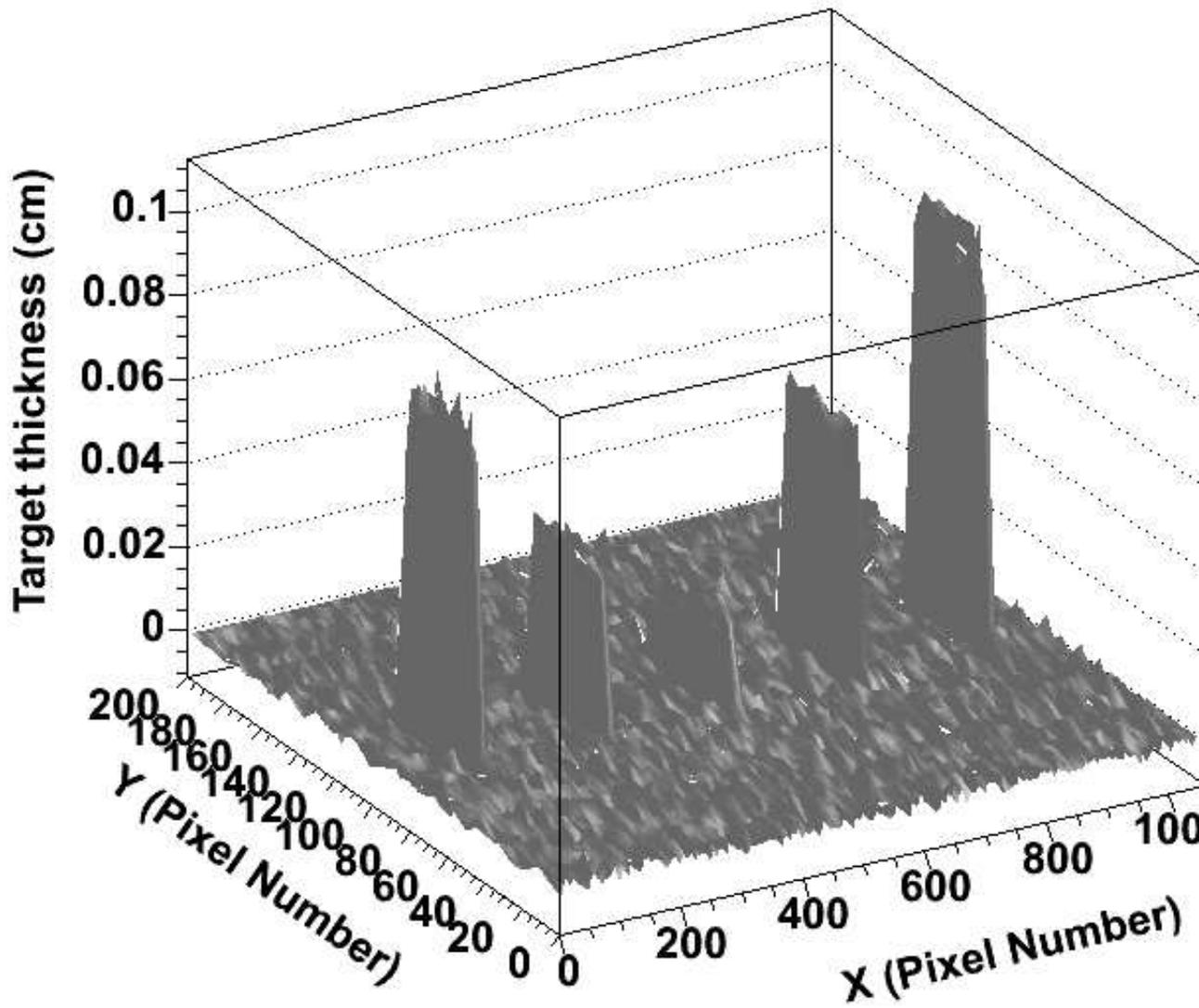}

\caption{Reconstructed 3D-image of the thicknesses for the {}``narrow''
beam. Original thicknesses are 0.08, 0.04, 0.02, 0.06 and 0.1 cm,
from left to right.}
\end{figure}

\begin{figure}[H]
\includegraphics[%
  scale=0.8]{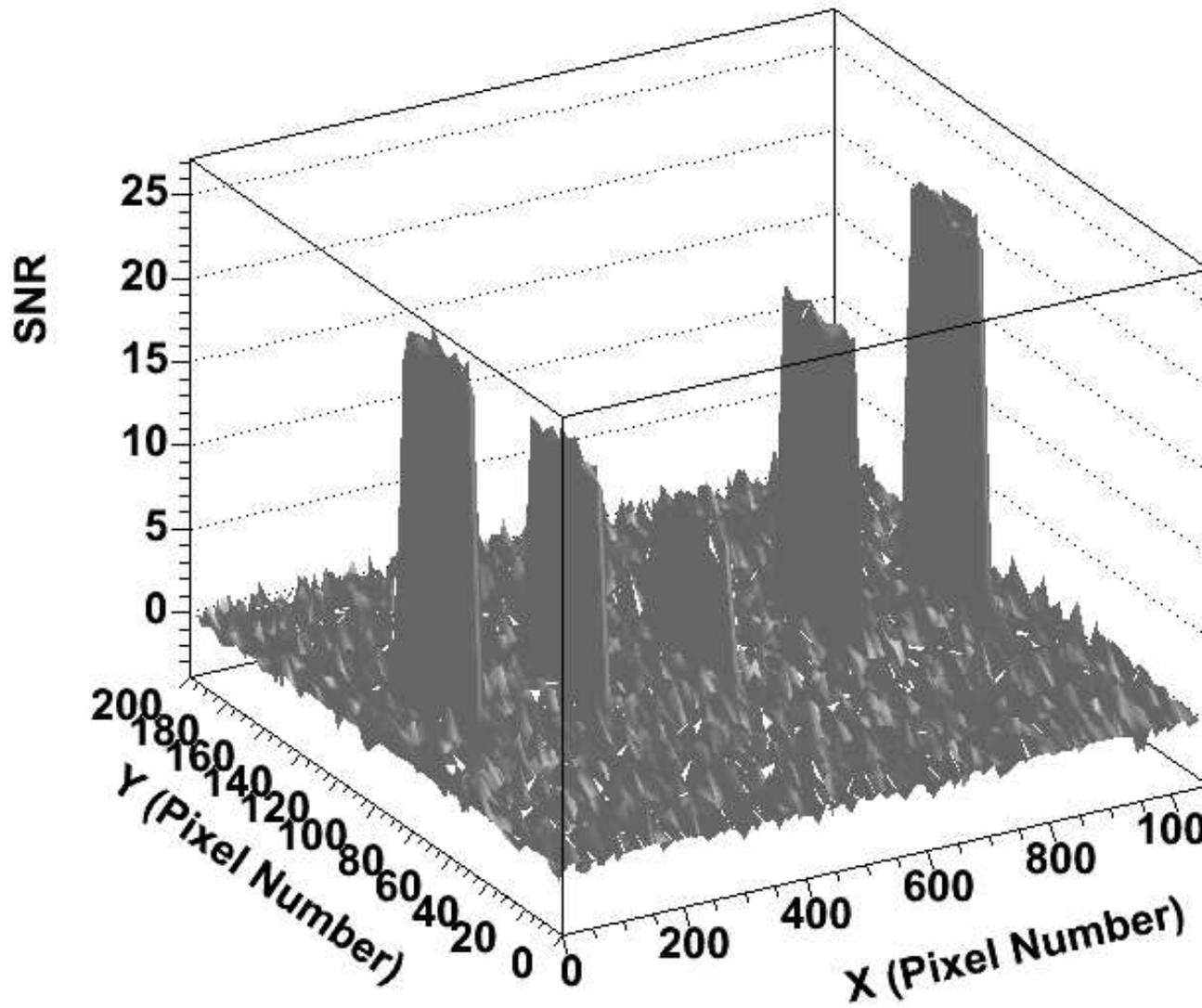}

\caption{Reconstructed 3D-image of the $SNR$ for the {}``narrow'' beam.}
\end{figure}

\begin{figure}[H]
\includegraphics[%
  scale=0.8]{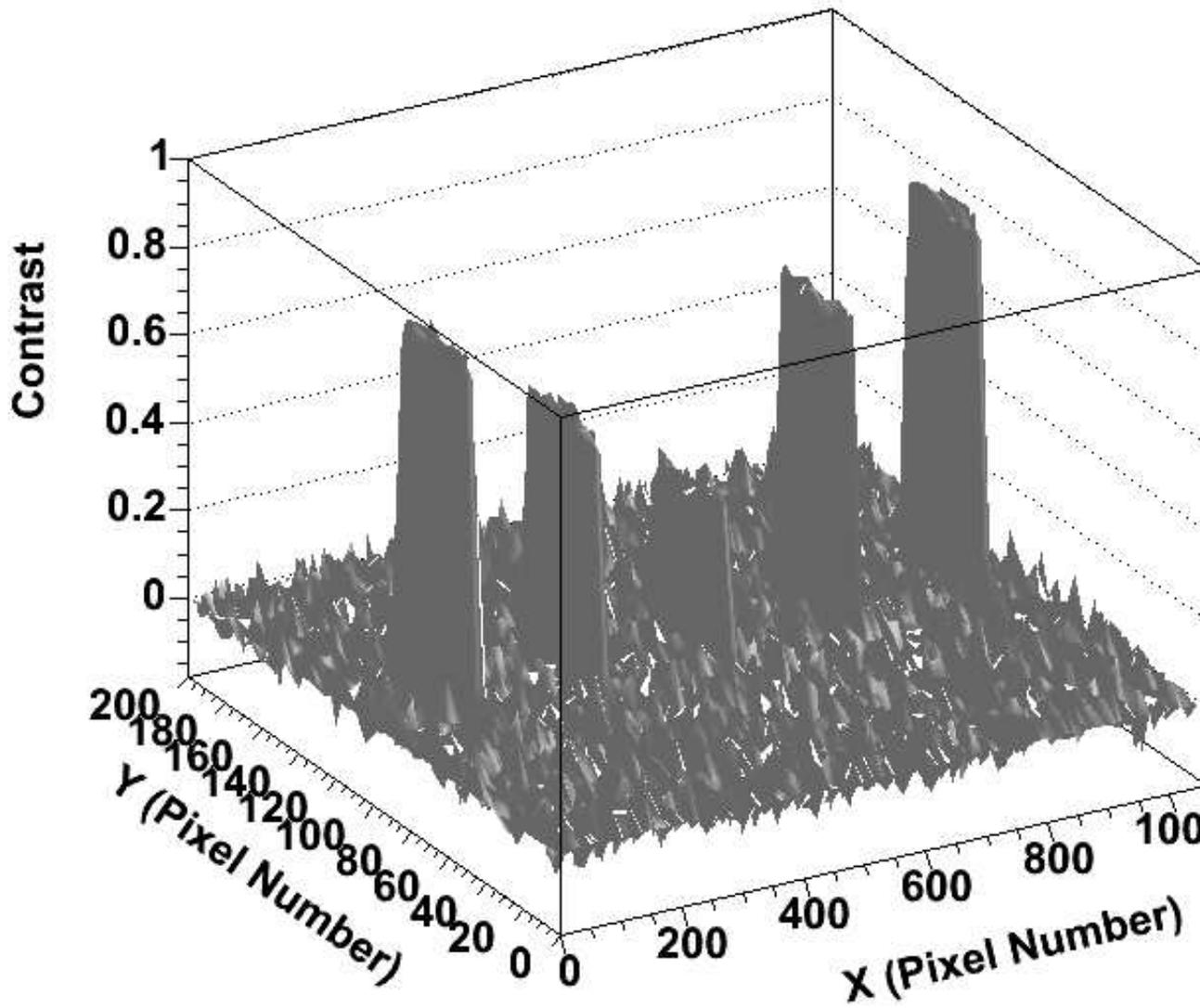}

\caption{Reconstructed 3D-image of the contrast for the {}``narrow'' beam
.}
\end{figure}

\begin{figure}[H]
\includegraphics[%
  scale=0.8]{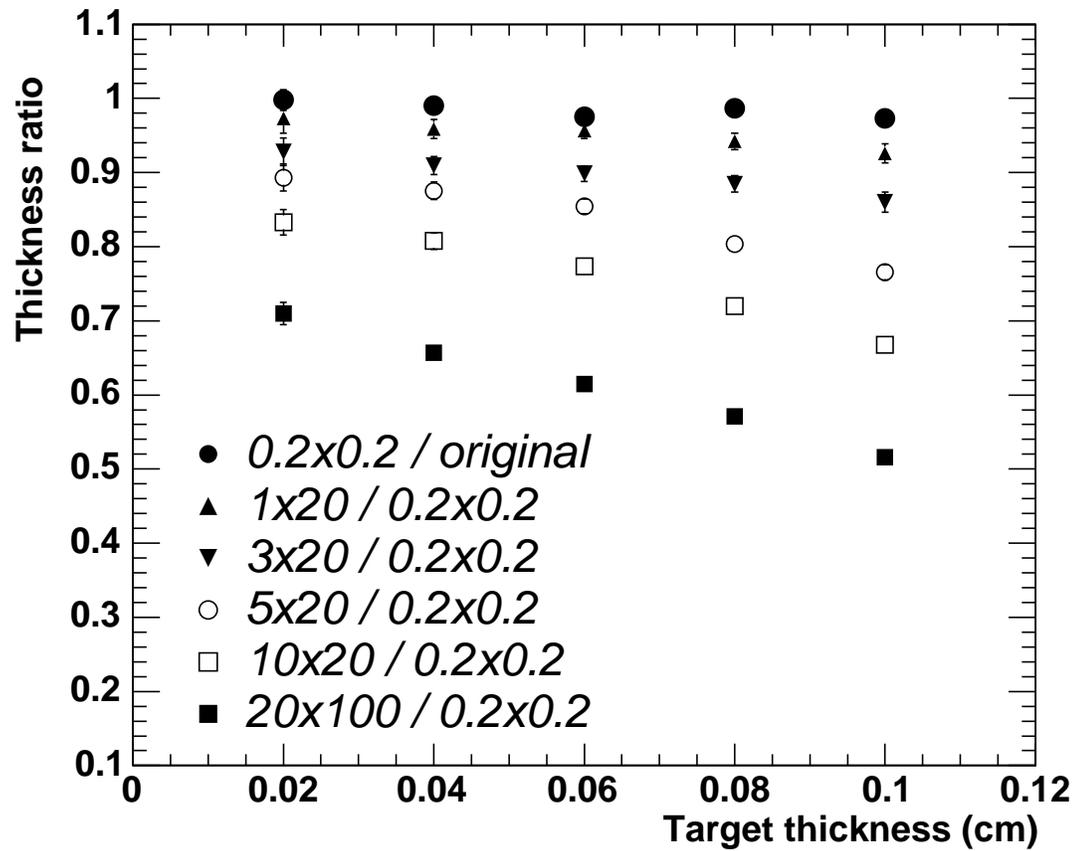}

\caption{The thickness ratio (with respect to the original thickness for the
narrow beam, and ratios for the rest ) in the simulated images using
the polynomial background definition, and its dependence on the $\mu C$
thicknesses for the different beams. Beam sizes are in $mm^{2}$}
\end{figure}

\begin{figure}[H]
\includegraphics[%
  scale=0.8]{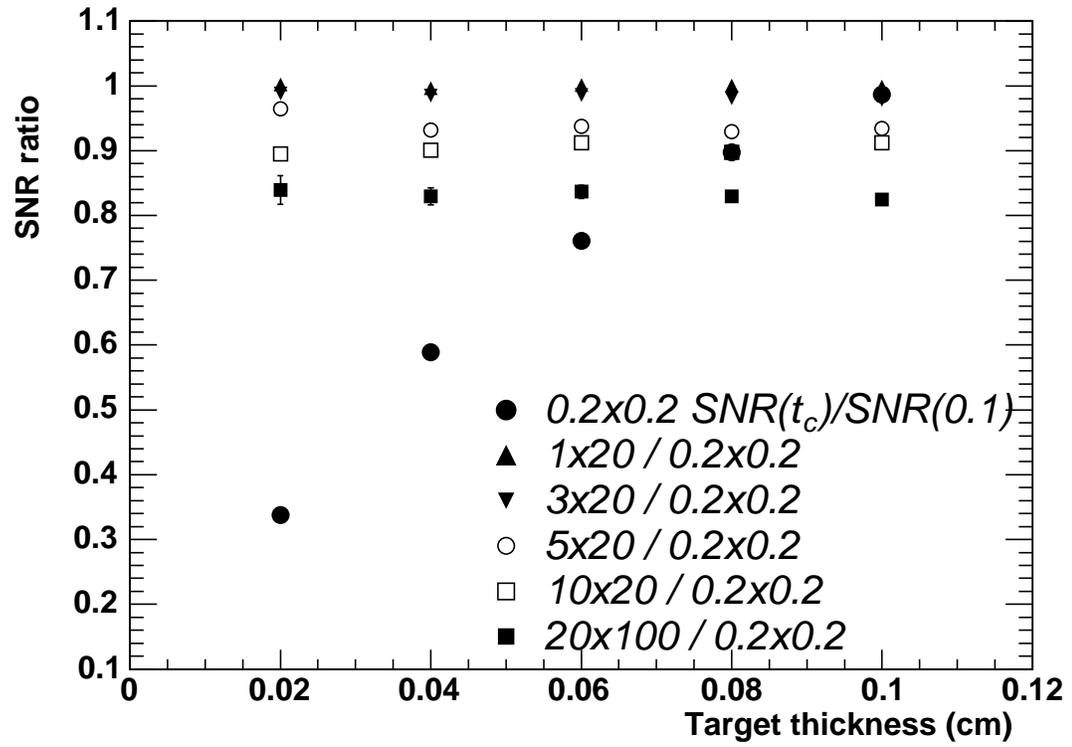}

\caption{The $SNR$ ratio in the simulated images using the polynomial background
definition, and its dependence on the $\mu C$ thicknesses for the
different beams. Beam sizes are in $mm^{2}$.}
\end{figure}

\begin{figure}[H]
\includegraphics[%
  scale=0.8]{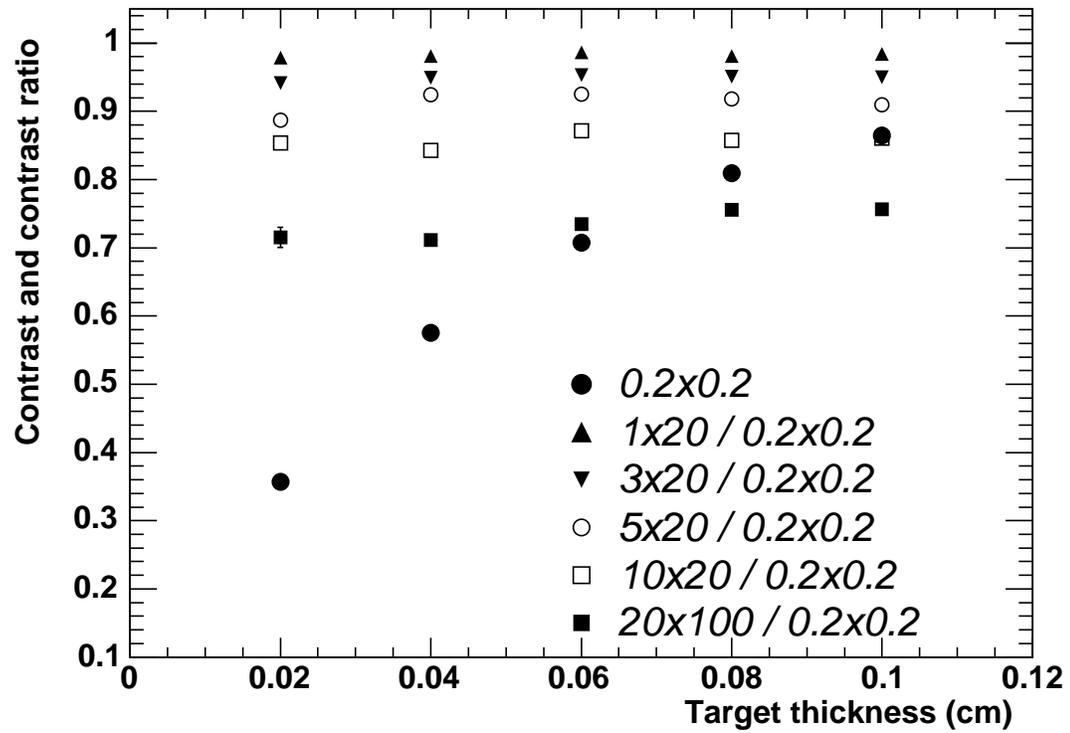}

\caption{The contrast (value for the {}``narrow'' beam and ratio value for
the rest) in the simulated images using the polynomial background
definition, and its dependence on the $\mu C$ thicknesses for the
different beams. Beam sizes are in $mm^{2}$.}
\end{figure}

\begin{figure}[H]
\includegraphics[%
  scale=0.8]{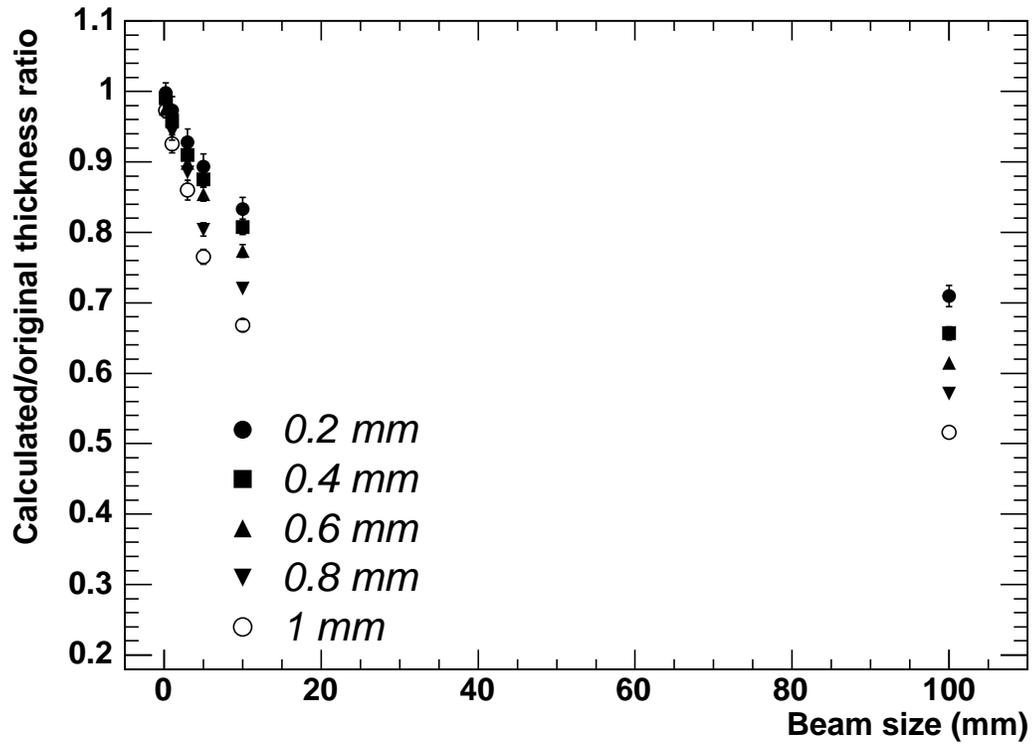}

\caption{Dependence of the thickness ratio (calculated/original) on the beam
size along the scanning direction (see Fig. 2). Different symbols
correspondent the original calcification thickness.}
\end{figure}

\end{document}